\documentclass[prl,twocolumn,showpacs,floatfix]{revtex4}
\usepackage{graphicx}
\usepackage{amssymb}

\begin{document}


\title{Magic angles and cross-hatching instability in hydrogel 
fracture}

\author{T. Baumberger, C. Caroli, D. Martina, and O. Ronsin}
\affiliation{INSP, UPMC Univ Paris 06, CNRS UMR 7588
140 rue de Lourmel, 75015 Paris France}

\date{\today}

\begin{abstract}
     The full 2D analysis of roughness profiles of fracture surfaces 
resulting from
     quasi-static crack propagation in gelatin gels reveals an
     original behavior characterized  by (i) strong
     anisotropy with maximum roughness at $V$-independent
     symmetry-preserving angles, (ii) a sub-critical instability 
     leading, below a  critical velocity, to a cross-hatched 
     regime due to straight macrosteps drifting at the same magic 
     angles and nucleated on crack-pinning network inhomogeneities. 
     Step height values are determined  by the width of the 
     strain-hardened zone, governed by the elastic crack blunting 
   characteristic of soft solids with breaking stresses much larger 
   that low strain  moduli.

\end{abstract}


\maketitle
Over the past two decades, considerable effort has been devoted to
characterizing and understanding the statistical properties of the
roughness of fracture surfaces in linear elastic disordered
materials. Investigation of a wide variety of systems, ranging from
brittle silica glass to ductile metallic alloys, has revealed the
ubiquity of wide roughness spectra exhibiting self-affine
characteristics on sizeable wavelength ranges. On the basis of
analysis of height-height correlation functions in a single
direction, E. Bouchaud {\it et al.} \cite{Bouchaud} first conjectured a fully
universal behavior summed up into a single Hurst exponent. The
discrepancies between the predictions of various subsequent theoretical
models have pointed toward the importance of also investigating
possible anisotropies of the scaling properties \cite{Ponson}. Work 
along this line
indeed reveals different scaling exponents along the crack
propagation direction and the (orthogonal) crack front one. From this,
Ponson {\it et al.} \cite{Bonamy} conclude that the self-affinity of 
the roughness
is described by a Family-Vicsek scaling, hence by a set of two
exponents. Within their new analysis, full universality is no longer the
case since they evidence the existence of at least two classes of
materials.

However, on the basis of their theoretical work, Bouchbinder
{\it et al.} \cite{Eran} have recently raised several new issues.
In particular,
they point to the necessity of a full 2D analysis of the
correlations and find, when reexamining some experimental data, that
a host of up to now unidentified exponents appear. They insist on the
need to focus on the effects of departures from linear elasticity in
the fracture process.

In this spirit we report here on the nature of surface morphologies
resulting from the fracture of gelatin, a highly compliant thermally
reversible hydrogel, in the strongly subsonic regime. We show that
these surfaces present very striking anisotropic features. Namely,
the rms roughness $R(\theta)$, measured along a direction at angle
$\theta$ from the propagation one $Ox$, exhibits two
symmetry-preserving maxima for $\theta = \pm \theta_{m}$. For gels
with a fixed gelatin concentration, this ``magic angle" $\theta_{m}$
is constant over more than one decade of both crack velocity $V$ and
solvent viscosity $\eta$. Moreover, below a critical,
$\eta$-dependent,
velocity $V_{c}$  a new  regime develops, characterized by a
cross-hatched (CH) morphology  analogous to those observed previously on a
covalently cross-linked hydrogel \cite{Seki,Tanaka} and on various
elastomers, swollen
or not \cite{Gent}. It is due to straight ``macroscopic" steps, here
of heights a few $100~\mu$m, emerging from the above-described
coexistent micrometric roughness. These steps are oriented at the
same magic angles $\pm \theta_{m}$ as those revealed by the
microroughness. Their growth results from a dynamical  instability 
which we show to be subcritical: for $V>V_{c}$, while the 
inhomogeneities of the gel network are not strong enough to trigger 
step nucleation, we are able to force  it with the help of externally 
applied local stress perturbations. 

These results  lead us to propose the
existence of a new class of morphologies, characteristic of highly
compliant random polymer networks with breaking stresses larger than 
their small strain Young modulus,  in which fracture implies elastic 
crack blunting \cite{Hui} and strong elastic
non-linearities at work in the near-tip region.

\begin{figure*}[ht]
\begin{center}
\includegraphics{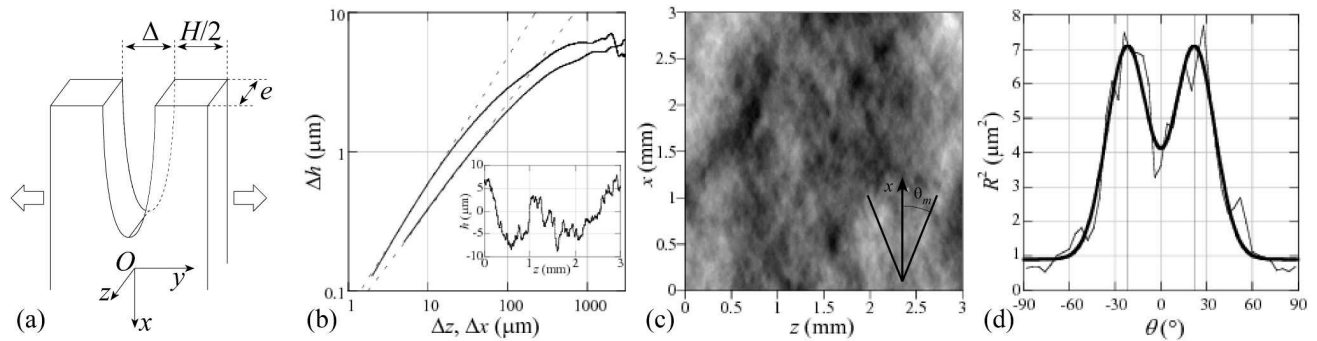}
\caption{\label{fig1} Roughness characteristics for a 5 wt\% 
 gelatin/water gel for crack velocity $V = 13.5$ mm/s.
 (a) Experimental geometry. 
Sample dimensions~: $H = 30$ mm,  $e = 10$ mm, length $L =
300$ mm. (b) Height-height correlation functions $\Delta
 h_{x}(\Delta x)$ (lower curve) and $\Delta h_{z}(\Delta z)$ (upper 
 curve). Inset : a typical profile along $Oz$. (c) $3\times 3$
mm$^{2}$ map of
heights $h(x, z)$. The gray scale ranges over $27\,\mu$m. The oblique
lines  correspond to peak positions in fig. 1.d. (d) Anisotropic squared
roughness $R^2$ {\it vs.} angle $\theta$ measured from crack
propagation direction. The dark line is the best fit with a sum of
two gaussians with an offset baseline. }

\end{center}
\end{figure*}

\paragraph {Experimental} --- Our fracture surfaces result from mode-I
cracks propagating along the mid-plane of long, thin gelatin slabs
(Fig. 1.a.), as described in detail in reference \cite{Nature}. A
displacement $\Delta$ of the rigid grips is imposed along $Oy$ to the notched
plate. After an initial transient, the crack reaches a
steady regime at a velocity $V$ which increases with the ``energy
release rate" $\mathcal G$ imposed by the opening $\Delta$. We found
that, in this slow, subsonic regime ($V<30$ mm/s):
\begin{equation}
    \label{eq:G}
 \mathcal G=\mathcal G_{0}+\Gamma\eta V
 \end{equation}
 with $\mathcal G_{0}\simeq
2.5$~J.m$^{-2}$, $\Gamma\simeq 10^{6}$, and were able to assign this
strong $V$-dependence to the fact that, in such a reversible gel,
fracture occurs via scissionless chain pull-out, the high dissipation
being due to viscous drag \cite{Nature,Nous}.

\begin{table}[ht]
\begin{tabular*}{8.6cm}{@{\extracolsep{\fill}}ccccc}
\hline
\hline
tip velocity (mm.s$^{-1}$) & 1.16 & 5.9 & 13.5 & 23.3\\
\hline
crack opening $\Delta$ (mm) & 5 & 7 & 9 & 11\\
\hline
rms roughness ($\mu$m) & 2.5 & 3.1 & 4.1 & 5.3\\
\hline
angle $\theta_{m}$ ($^{\circ}$) & 18 & 17 & 17 & 22\\
\hline
\hline
\end{tabular*}
\caption{}
\end{table}
The results reported here were obtained on gels with a $5$ wt.\%
content of gelatin in the solvent (pure water except when otherwise
specified). The gel low-strain shear modulus $\mu = 3.5$~kPa
(transverse sound speed $\sim 1.9$~m/s), is due to the entropic
elasticity of the random polymer network, of average mesh size
$\xi\sim 10$~nm. Note that, owing to this huge compliance, even very
slow cracks ($V\sim 100~\mu$m/s) demand large openings $\Delta$
(Table I),
typically
$\gtrsim 3$ mm. Post-mortem
replicas of crack surfaces, obtained by UV curing of a thin layer of
glue, are scanned with the help of a mechanical profilometer (stylus
tip radius $2\,\mu$m, out of plane resolution $0.1$~nm). Our maps
$h(x,z)$ of heights above mean plane correspond to $3\times 3$
mm$^{2}$ scans. Each one consists of a series of 600 lines parallel
to the front direction $Oz$, each line containing 1500 data points.

\paragraph{The microrough morphology} --- For $V\gtrsim 300~\mu$m/s,
the fracture surfaces are flat on the macroscale, yet
clearly not mirror-like. As expected with such a material, elastic
modulus as well as toughness fluctuations due to the randomness of
the network result in in- and out-of-plane excursions of the crack
front line, the profile roughness being the imprint of the latter
ones. A typical map is shown in figure 1.b. For this crack velocity $V
= 13.5$ mm/s, we find that the global rms roughness $\bar R =
4.1\,\mu$m \cite{foot}.
Following Ponson {\it et al.} \cite{Bonamy}, we first
characterize height correlations along  the front
($Oz$) i.e. we compute~:
$\Delta h_{z}(\Delta z) = \left\langle\left\lbrack h(z+\Delta z,
x)-h(z, x)\right\rbrack\right\rangle_{z, x}^{1/2}$ and its counterpart
$\Delta h_{x}(\Delta x)$
along the propagation direction ($Ox$). As indicated by figure 1.b. the
microroughness spectrum is anisotropic and of the broad band type.
However, at variance with the findings of \cite{Bonamy}, a simple
power law fit (see dashed lines) is, here, clearly unconvincing. The
discrepancy leads us, following Bouchbinder {\it et al.}
\cite{Eran}, to try and
exploit more extensively the full 2D information. Indeed, naked-eye
inspection under grazing incidence, as well as the aspect of height
maps (Fig. 1.c.), strongly suggest the presence of preferred oblique
directions. This suspicion we have confirmed by computing the
angle-dependent squared roughness $R^2(\theta)$, obtained by integrating the
power spectrum $\vert\tilde h({\bf q})\vert^{2}$ over an angular
sector $d\theta = 4^{\circ}$ about angle $\theta + \pi/2$ (in $\bf
q$-space) and in the wavelength window $2\,\mu$m--$1$ mm. We find in
all cases that $R^2(\theta)$ exhibits two marked symmetric peaks at
$\theta =\pm\theta_{m}$ (see figure 1.d. for which $\theta_{m} =
22^{\circ}$). Moreover (Table I), $\theta_{m}$ does not show any significant
variation over the whole $V$-range ($V_{max}/V_{min}\sim 20$). In
addition, a 10-fold increase of solvent viscosity $\eta$, obtained by
using a $40/60$ water/glycerol mixture, does not either affect its
value.

Finally (see Table I), the full 2D-averaged rms roughness $\bar R$ 
clearly increases 
with the imposed crack opening $\Delta$, hence with $V$.

\paragraph{The crosshatched (CH) morphology} --- Starting from
the microrough regime, we perform a step-by-step decrease of $\Delta$,
thus slowing down crack propagation. Below a critical velocity $V_{c}
= 350 \pm 20\,\mu$m/s we observe on the fracture surfaces the emergence,
at apparently random locations across the slab width, of oblique
straight line defects (Fig. 2.a.). Upon further $V$ decrease, these
lines proliferate into a scale-like pattern, akin to the CH
morphologies found by Tanaka {\it et al.} \cite{Seki} on a covalently
cross-linked gel and by Gent {\it et al.} \cite{Gent} on elastomers. 

Note that, since these observations pertain to the quasistatic regime, front 
wave propagation \cite{Fineberg} is irrelevant here.

\begin{figure}[ht]
\begin{center}
\includegraphics{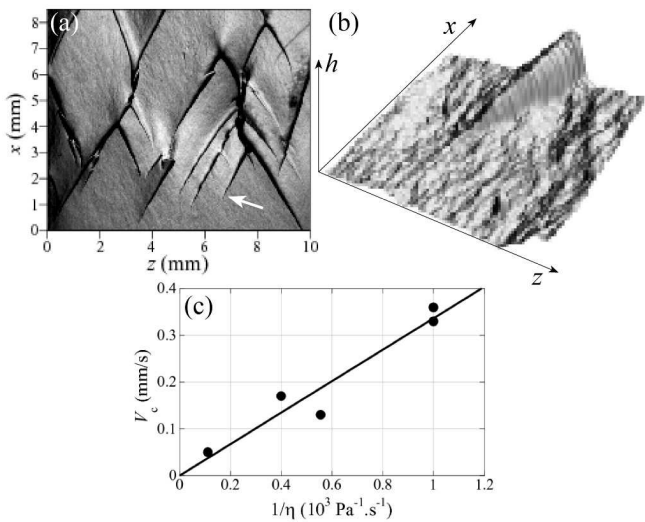}
\caption{\label{fig2}Emergence of the CH regime. (a) A typical CH 
pattern close below $V_{c}$. The crack propagates 
upward. (b) 3D profile (400$\times$400~$\mu$m$^2$,
15~$\mu$m peak to peak)  the
 emergence of a macrostep (arrow on Fig. 2.a.) out of the 
 microroughness. (c) Critical velocity $V_{c}$ vs. inverse viscosity 
 for 5\% gelatin gels in water/glycerol solvents. The line is the 
 best linear fit.}

\end{center}
\end{figure}

Profile analysis (Fig. 2.b.) shows that these lines correspond to steps which 
emerge via  self-amplification of the microroughness, with which they
coexist. Their height saturates, after a traveling distance of
several millimeters, to values on the order of $200\,\mu$m. No change of their
drift angle is discernable as they grow, so that they propagate at
the magic angle $\theta_{m}$ of the microroughness. We find it to
remain
$V$-independent over the full $V<V_{c}$ explored range.

The large $\Delta$ values make it possible to image the crack 
surfaces in the tip region. Figures  3.a--d depict the emergence and 
growth of a defect, in a frame attached to the moving crack front. It 
is seen to drift along the front direction $Oz$ at a constant velocity 
$V_{d}$, as expected from the straightness of CH lines. So, as we have checked from \emph{e.g.} Fig. 3.b-d., $V_{d} 
= V \tan\theta_{m}$. Fig. 3.e  corresponds to the fully 
developed defect, schematized on Fig. 3.f. It
shows that the crack front has become 
discontinuous, and composed of two semi-cracks $A_{0}A$, $B_{0}B$. 
As it moves along $Ox$, $A_{0}A$ creates two ``lips" $\alpha_{I}$, 
$\alpha_{II}$. In the vicinity of its tip, the second 
semi-crack is delayed with respect to $A_{0}A$, so that its end point 
$B$ lies on the $\alpha_{II}$ lip. The defect topology thus involves 
a  material overhang $ABC_{I}D_{I}$. Note that this topology is the 
equivalent, for a widely open crack, of that identified in 
\cite{Seki} and \cite{Gent}. As the front proceeds, the fold 
$A_{I}C_{I}D_{I}B_{I}$ ultimately collapses, leaving two 
complementary steps on the crack faces. 

Finally, in order to shed light on the role of solvent viscosity $\eta$, 
we have studied gels made with water-glycerol mixtures. We find that 
the CH morphology persists but that the critical velocity $V_{c}$ 
scales as $1/\eta$ (Fig. 2.c). However, the magic angle $\theta_{m}$ is 
$\eta$-independent. 

\paragraph{Discussion} --- The above phenomenology presents several
striking discrepancies with the known morphologies,  pertaining to
metallic alloys, silica glasses, quasi-crystals, mortar, wood and
sandstone \cite{Ponson}. Namely, none of these materials has been reported to
exhibit (i) preferential anisotropy directions (ii) a CH regime (iii)
a noticeable increase of $\bar R$ with $V$ in the quasistatic regime.
On the other hand, these  a priori puzzling features are not
restricted to gelatin gels. Indeed, as already mentioned, the CH
regime has been documented for a chemical hydrogel and for
elastomers. Moreover, using figure 5 of reference \cite{Tanaka}, one easily
checks that, in their chemical gel as well, the CH drift angle is constant
($\theta_{m}\simeq 43^\circ$) over a 40-fold increase of $V$ up to a
critical $V_{c}$ where the CH regime disappears.

\begin{figure}[ht]
\begin{center}
\includegraphics{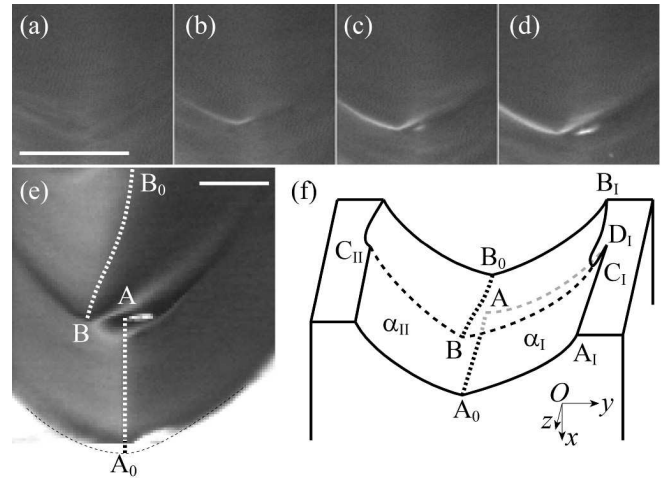}
\caption{\label{fig3}Emergence and growth of a CH defect as viewed  at $60^{\circ}$ 
from  $Ox$ in the 
$Oxz$ plane. White bar length : 200 $\mu$m (a)--(d) Snapshots at constant time intervals $\Delta t 
= 0.4 s$ in the moving tip frame. The defect (bright line) grows and 
drifts at constant velocity. (e) Fully developed defect. The wite 
dashed lines mark the 
discontinuous front line. (f) Sketch of the defect topology.} 
\end{center}
\end{figure}

The question then arises of which material features may be
responsible for this new class of behavior. 
In view of the dependence of $V_{c}$ on solvent viscosity, one might 
be tempted to invoke poroelasticity. However Gent et al. have observed 
CH morphologies as well 
with dry elastomers as with a network swollen (ration 2.6) with oil. 
Hence the presence of solvent is not necessary for this regime to 
exist. 

This apparent paradox has led us to investigate the nature of 
the transition at $V_{c}$. The fact that (i) the microroughness does 
not show any trend towards divergence when approaching $V_{c}$ from 
above, (ii) the height $\delta$ of CH steps does not vanish as
 $V\to V_{c}^-$, prompts that the CH instability is 
discontinuous, i.e. involves nucleation. If so, it should be possible 
to trigger  CH steps above $V_{c}$ with the help of 
large enough external perturbations. Indeed, in a large range of 
$V>V_{c}$, inducing a mixed 
(II+III) mode stress perturbation by pinching the gel slab 
off mid-plane, slightly 
ahead of the tip, results in the nucleation of a step that propagates 
through the sample. 

So, $V_{c}$ appears as a threshold below which the structural noise 
due the randomness of the gel network is not strong enough for the 
step nucleation barrier to be overcome. This, together with the fact 
that $\eta V_{c}$ is a constant for a given network structure, suggests 
the following tentative picture: when the crack front meets a tougher 
zone (due to a local decrease $\delta \xi$ of meshsize $\xi$) of 
spatial extent $\gtrsim d_{act}$, with $d_{act}\approx 100$ nm 
 the size of the process zone\cite{Nature}, it slows down locally, 
generating a wall which stretches as the crack lips open. We 
postulate that step nucleation demands that the tough zone pins the 
front down to a full stop, i.e. that the front velocity fluctuation 
$\delta V = -V_{c}$. According to the fracture model developed in 
refs. \cite{Nature}, \cite{Nous}, both $\mathcal G_{0}$ and $\Gamma$ 
in eq.(\ref{eq:G}) scale as $\xi^{-2}$. In the experiment, $\mathcal 
G$ is kept constant via the constancy of $\Delta$. So, 
$\eta\delta V = (\mathcal G /\Gamma)\delta\xi^{2}/\xi^{2}$. From 
this, we can estimate the maximum relative amplitude $\epsilon_{m}$ of mesh size 
fluctuations on spatial scales $\gtrsim d_{act}$ in our system as 
$\epsilon_{m} \approx \Gamma/2\mathcal G_{0}\eta V_{c} \approx 7\% $. The 
smallness of $\epsilon_{m}$ justifies our approximating $\mathcal G$ by 
$\mathcal G_{0}$ in this estimate and explains the quasi-constancy of 
$\eta V_{c}$. Such an order of magnitude  is statically 
plausible, and at the same time compatible with the sizeable large-angle 
light scattering exhibited by our gels. 

On the basis of this picture we interpret step nucleation as 
resulting from  the  pinning   of a small part of the crack line. Due 
to local asymmetries, one of the resulting semi-cracks takes the 
lead, the other one getting delayed while skirting around the tough 
zone via an excursion towards the easier lip, thus nucleating the 
overhang. As cracking proceeds its endpoint $B$ is gradually advected 
away by the opening of the $A_{0}A$ crack (Fig. 3.f), until it gets 
far enough from $A_{0}A$ for tensile stresses along $\alpha_{II}$ to 
become negligible.  

Now, our gelatin gels share with all the other soft solids exhibiting 
CH two mechanical characteristics: (i) the breaking stress is much 
larger than the low strain Young modulus $E$ (ii) they strain-harden
considerably at large deformation levels where 
polymer segments are stretched quasi-taut\cite{GentNL}. 
Hui et al. \cite{Hui} have 
shown that (i) entails elastic crack blunting while, due to (ii) 
stress concentration is restricted to a strain-hardened region of 
thickness $\sim \mathcal G/E$ along $Oy$. This allows us to predict 
that outward advection of $B$ stops when the distance from $B$ to 
$A_{0}A$, i.e. the step height $\delta$ becomes of order $\mathcal 
G_{0}/E$. With $\mathcal G_{0}\approx 2.5$ J.m$^{-2}$, $E\approx 
10$ kPa, we get $\delta\approx 150\,\mu$m, in excellent agreement with 
the experimental value. For slow cracks in elastomers as studied in 
\cite{Gent}, $\mathcal G_{0}$ and the relaxed $E$ are respectively of 
order a few 10 J.m$^{-2}$ and a few MPa, leading to step heights 
$\delta$ in the $10\,\mu$m range. This agrees with Gent's findings, 
lending further support to our picture.  

Finally, let us reconsider the defect shape as depicted on 
Fig. 3.f. In its end region close to $A$ (i.e. for $z_{A}<z<z_{B}$), 
the lower semi-crack 
connects the stretched lip $\alpha_{I}$ with the ceiling of the 
fold, in which stresses are, to a large extent, relaxed by the 
$B_{0}B$ cut. This is consistent with the fact that, as cracking 
proceeds, $B_{0}B$ which cuts into the unscreened $\alpha_{II}$ lip, 
elongates ($B$ drifts) towards $z>0$ while $A_{0}A$ regresses. This screening effect 
of the fold geometry explains that the defect drifts along the front 
in a direction defined by the initial symmetry-breaking choice which 
has led $B_{0}$ towards one or the other crack lip. 

In the context of the above description, the existence of a 
preferred orientation in the micro-roughness could be attributed to 
aborted attempts at fold development. However, why $\theta_{m}$ is 
independent of both $V$ and the amplitude of out-of-plane excursions 
remains at this stage a fully open question. 

In conclusion we propose that the phenomenology brought to light and
analyzed here
is characteristic, beyond the case of gelatin gels, of a new class
of morphologies of fracture
surfaces. The scenario which we propose to describe the 
cross-hatching instability implies that this class is characteristic 
of soft, tough, elastic materials which exhibit elastic 
crack-blunting and strong, strain-hardening, elastic 
non-linearities.\\

\begin{acknowledgments}
     We are grateful to K. Sekimoto for stimulating interactions along
     the course of this work. We also thank M. Adda-Bedia, A. Argon,
     D. Bonamy, H. Brown, V. Lazarus for helpful discussions.
\end{acknowledgments}


\begin{thebibliography}{100}
\bibitem{foot} Computation of the real surface area shows it to
differ by less than 1\% from the apparent (projected) one, so that the
corresponding correction to the fracture energy is negligible.
\end{thebibliography}

\end{document}